\documentstyle[12pt,epsfig]{article}

\newcommand{\Bach}{{\cal B}}
\newcommand{\Ach}{{\cal A}}

\begin{document}

\pagestyle{empty}

\title{Risk-return arguments applied to options
with trading costs}
\author{ Erik Aurell$^{1,2}$, Karol \.Zyczkowski$^3$ \\
         $^1$ Dept. of Mathematics, Stockholm University, S-106 91 Stockholm, Sweden \\
         $^2$ PDC/KTH, S-100 44 Stockholm, Sweden \\
	 $^3$ Dept. of Physics, Jagiellonian University,  PL-30 057 Krak\'ow,
		Poland}
\maketitle

\begin{abstract}
We study the problem of option pricing and hedging strategies
within the frame-work of risk-return arguments. An economic
agent is described by a utility function that depends on
profit (an expected value) and risk (a variance).
In the ideal case without transaction costs the optimal strategy
for any given agent is found as the explicit solution of a constrained
optimization problem. Transaction costs are taken into account
on a perturbative way.
A rational option price, in a world with only these
agents, is then determined by considering the points of view
of the buyer and the writer of the option.
Price and strategy are determined to first order in the transaction
costs.
\end{abstract}

\section{INTRODUCTION}
Options are financial contracts made out between two
economic agents called the {\it writer} and
the {\it buyer}. The content of such a contract is that
it gives
the buyer the optional right to buy from
the writer a unit of some
commodity at some time in the future at 
a determined price. Options differ mainly as to the type
of underlying commodity (stock, stock indices, 
foreign currency, etc.), if the expiry time
is fixed or may be chosen by the option buyer
(European or American), and on the form of the pay-out function.
Option pricing theory is generally regarded
as one of the corner-stones
of modern mathematical finance, 
for standard text-book treatments, see
\cite{Duffie,CoxRubinstein,Hull}. The outcomes of these
theories are normative prescriptions of {\it option prices} 
and {\it hedging strategies}, the latter being portfolios of the
underlying commodity to be held in conjunction with the option.
The theory has generally been developed for the class of
complete markets, the two main examples being
the log-Brownian continuous-time model of
Black and Scholes\cite{BlackScholes}, and
the dichotomic discrete-time model of Cox, Ross
and Rubinstein\cite{CoxRossRubinstein}.

Option pricing in incomplete markets have been much
less well developed, partly because the Black-Scholes
and Cox-Ross-Rubinstein theories already yield quite reasonable
estimates of  observed market
prices, partly because there is no agreed-upon procedure
to find the price in this more general situation.
In economic terms this is is expressed as the price
being contingent on individual investor attitudes.

The perhaps simplest assumption about investor attitudes
is that they
can be described only by their preferred trade-offs of 
risk versus return on investment.
If the risk is measured
by the standard deviation,
then we are effectively looking at the option problem
in the spirit
of the Markowitz portfolio theory\cite{Markowitz}.
Even though this is quite a simplification, there
remains (at least) one parameter describing the
risk-aversiveness of an individual investor, which
appears as the parameter $\lambda$ below.
To have predictive power we must be able to say something
about a price on the market without assuming, say,
that all investors can be described by the same parameter
$\lambda$.
One recalls that in the Capital Asset Pricing Model
one is able to derive a relation between returns and
(diversified) risks on the market\cite{Sharpe,Litner}.
The particular relation (the slope of the Capital Market
Line) depends on the set of investment opportunities
and investor attitudes, but once it has been established, it
holds for all investors. In short, one would like to get
out something similar from a mean-variance approach to
option pricing.
 
One way to proceed is to postulate that the hedging problem
is solved by minimizing risk unconditionally.
This approach has been put forward by
several authors, notably by Schweizer\cite{Schweizer}
and Bouchaud \& Sornette \cite{BouchaudSornette}.
In the mean-variance language it directly
corresponds to all investors
being very risk-aversive, characterized by very large
values of $\lambda$. If all investors share this attitude,
and trade with one another, then the expected profit from each
trade has to be zero, and this argument fixes the price.

When presented this way the risk-minimizing prescription has
problems: Why all operators have to be very risk-aversive?
And if they are very risk-aversive, and make on the average
zero net profit on an option trade, then why would they bother
to engage in it?
However, it turns out that the special role of risk-minimizing
hedging can be derived from a mean-variance approach in a
different manner\cite{AurellZyczkowski}.
The main objective of this present paper is to present this result, 
including transaction costs in the calculations.

In any reasonable approach no
economic agent should be prepared to sell an option more cheaply
than the price he would be prepared to pay for it. 
There may however be some normative agent who is prepared to buy
and sell at one and the same price. If that is the case,
and all other agents buy at
lower and sell at higher prices, then this price is in fact
a possible market price.
This is what happens in the mean-variance
approach without transaction costs.
It is more appealing than the straight-forward risk-minimizing
procedure is that the normative agent is neither infinitely risk-aversive
nor infinitely risk-willing, but something in between.
The case of agents still more risk-willing requires a discussion
outside the scope of this paper, see \cite{AurellZyczkowski}, but
does not change the argument.
The price proposed by
the normative agent is such, that if he used the risk-minimizing
strategy his expected profit would be zero. His price therefore
agrees with that of the risk-minimizing prescription of
Schweizer and Bouchaud-Sornette.
Ultimately this is a consequence
of the variance being a quadratic functional of the strategy,
and it would not hold if we adopt another measure of risk.
The strategy actually used by the normative agent is however
not the risk-minimizing strategy, It can best be described as the
the risk-minimizing strategy plus a component of pure investment
in stock, unrelated to the option. 
In this way we can rederive the {\it price} of the
risk-minimizing approach, but we do not imply that any 
rational agent actually uses that {\it strategy}, except as
part of a larger portfolio.

The paper is organized as follows: in section~\ref{s:no-transaction-costs}
we set up the problem and
rederive the results on mean-variance
optimal portfolios of \cite{AurellZyczkowski}.
In~section~\ref{s:transaction-costs}, which contains the hard new
results of the paper, we introduce transaction costs,
and see how they modify the price and the strategy.
In~section~\ref{s:buyer-seller}, included for completeness, we look at 
the buyer's and the seller's side of the market, and summarize the results.

\section{RISK-RETURN WITHOUT TRANSACTION COSTS}
\label{s:no-transaction-costs}
We now specialize the discussion to a European
Call Option contracted at time $0$ to expire at
time $T$, with strike price $S_c$. 
Time is taken discrete
in units of $\tau$, such that $T=N\tau$.  
For convenience we refer to the underlying commodity as
stock, which comes in units of shares. The price of the
share is assumed to generated by a multiplicative random
processes, such that $S_{i+1}=u_{i}S_{i}$, where the
$u_i$'s are independent and identically distributed random
variables with finite variance.
An important parameter is $\mu$, the expected excess
return on the share compared to a risk-less investment.
Let us assume that a risk-less investment increases
in value from $1$ to $r$ over one discrete trading period,
then
\begin{equation}
\mu = <(u_i-r)>
\label{eq:mu}
\end{equation}
In realistic market models one would
expect $\mu$ to be larger than zero.
The incremental wealth of an option writer that sells the option
for $C$ and uses the strategy to
keep $\phi_i(S_i)$ shares against the option at time $t=i\tau$,
if the realized share price is $S_i$, is, in the absence
of trading costs,
\begin{equation}
\Delta W = Cr^{N} +  \sum_{i=0}^{N-1}(u_i-r) r^{N-i-1}S_i \phi_i(S_i) - \{S_T-S_c\}^+
\label{eq:DeltaW}
\end{equation}
We follow here the notation of
\cite{BouchaudSornette}, to which we
refer for a discussion and motivation of (\ref{eq:DeltaW}),
see also \cite{AurellZyczkowski,AurellSimdyankin}.

The gain and the risk are the expected value and the variance of
the incremental wealth, with the price substracted, i.e.
\begin{eqnarray}
M[\phi]\, &\, =\,& <\Delta W - Cr^{N}>  \label{drift_appendix} \\
R[\phi] \,&\,=\,&<(\Delta W - Cr^{N})^2>\, -\, \left(<\Delta W - Cr^{N}>\right)^2
\label{risk_appendix}
\end{eqnarray}
The profit, the expected value of $\Delta W$ in (\ref{eq:DeltaW}), is equal
to the gain plus $Cr^{N}$.

$M$ is a linear and $R$ is a quadratic functional of $\phi$.
If we perform 
the minimization of $R$ with $M$ constrained to the
value $m$, it is clear that this $R$ must
be a quadratic polynomial in $m$.
To get the explicit coefficients of that polynomial 
we introduce the vector-valued
set of functions given by
\begin{eqnarray}
F_i(S_i)& =& <\{S_T-S_c\}^+ (u_i-r)>_{S_i} ,
\end{eqnarray}
the auxiliary variable
\begin{equation}
\tilde \psi_i(S_i) = 
P(S_i|S_0) r^{N-i-1}S_i \phi_i(S_i) ,
\label{tilde-psi-definition}
\end{equation}
and the matrix
\begin{equation}
K_{ij}(S_i,S_j) = \frac{P(S_i,S_j|S_0)}{P(S_i|S_0)P(S_j|S_0)}
<(u_i-r)(u_j-r)>_{S_i,S_j}\,  - \, \mu^2,
\label{K-definition}
\end{equation}
where, naturally, $P(S_i|S_0)$ is the conditional probability
that the share price equals $S_i$ at time $i$, given that it
was initially $S_0$. Likewise $P(S_i,S_j|S_0)$ is the joint probability
that the price equals $S_i$ at time $i$ and $S_j$ at time $j$,
conditioned by having been initially  $S_0$.
The diagonal elements of $K$ can be written a little more simply
as
\begin{equation}
K_{ii}(S_i,S_i) = \frac{1}{P(S_i|S_0)} <(u_i-r)^2>_{S_i} \,  - \, \mu^2.
\label{K-diagonal-definition}
\end{equation}
In the special case when $\mu$ is equal to zero
$K$ is diagonal.
The gain and the risk can now be written as ordinary
scalar products involving $\tilde\psi$, $K$ and $F$:
\begin{eqnarray}
M &=& \Bach + \mu(\tilde\psi\cdot {\bf 1})
\label{drift_appendix_simplified}  \\
R &=& R_c -2( {\bf F \cdot  \tilde\psi} ) +2\mu\Bach ( {\bf \tilde\psi \cdot  1})
+ ( {\bf \tilde\psi} \cdot  K{\bf \tilde\psi})
\label{risk_appendix_simplified}
\end{eqnarray}
where the notation $\Bach$ stands for the average
$<\{ S_T-S_c\}^+>$, and $R_c$  the corresponding variance,
$<(\{ S_T-S_c\}^+)^2> - (<\{ S_T-S_c\}^+>)^2$.
The minimum risk at given gain $m$ is 
\begin{equation}
R[m] = \rho + \nu(m+\Ach)^2
\label{RofMansatz}
\end{equation}
with the following identification of the coefficients;
\begin{eqnarray}
\rho & =&  R_c - ({\bf F  }\cdot K^{-1}{\bf F} )
  +2 \Bach\mu ({\bf 1}\cdot K^{-1}{\bf F}) 
%\nonumber \\ &&\qquad\quad
 -\Bach^2\mu^2 ({\bf 1}\cdot K^{-1}{\bf 1})
 \label{rhodef} \\
\Ach & =& \Bach -\mu ({\bf 1}\cdot K^{-1}{\bf F}) +\Bach\mu^2 ({\bf 1}\cdot K^{-1}{\bf 1})
\label{Mhatdef} \\
\nu & =& \frac{1}
{\mu^2({\bf 1}\cdot K^{-1}{\bf 1})} \label{nudef}
\end{eqnarray}
The optimal strategy is given by
\begin{eqnarray}
\tilde\psi_i(S_i;m) &=& (K^{-1}{\bf  F})_i(S_i) +
%\nonumber \\ &&
{{m + \Bach -\mu({\bf 1}\cdot K^{-1} {\bf F})}\over{\mu
({\bf 1}\cdot K^{-1}{\bf 1}) }}
(K^{-1}{\bf 1})_i(S_i).
\label{GofM}
\end{eqnarray}
The value of $m$ such that the lowest risk is attained is 
$m=-\Ach$. The risk-minimizing strategy is thus
\begin{eqnarray}
\tilde\psi_i(S_i;-\Ach) &=& (K^{-1}{\bf  F})_i(S_i) -\mu\Bach (K^{-1}{\bf 1})_i(S_i).
\label{GofMstar}
\end{eqnarray}
The expected profit when trading with the risk-minimizing strategy is
$Cr^N-\Ach$. If we adjust the price ($C$) such that the expected profit is zero,
we have
\begin{eqnarray}
C = r^{-N}\Ach
\label{Cdefinition}
\end{eqnarray}
where $\Ach$ is given in (\ref{Mhatdef}).
Equations~(\ref{GofMstar})
and~(\ref{Cdefinition}) summarize the
risk-minimizing approach to option
pricing and hedging without transaction costs.

\section{RISK-RETURN WITH TRANSACTION COSTS}
\label{s:transaction-costs}
We now assume that trading costs are present in the form
\begin{equation}
\Delta W_{costs} = -\gamma F[\phi]
\end{equation}
where $\gamma$ is a small parameter and $F$ is 
a positive functional
of the strategy $\phi$. For instance, $F$ could include 
proportional trading costs  when changing the portfolio
from $\phi_i$ to  $\phi_{i+1}$\cite{BouchaudSornette}.

We will assume that agents try to 
maximize a utility function of the following kind
\begin{equation}
U[\phi,\gamma;\lambda] = M[\phi,\gamma]+Cr^N -\lambda\sqrt{R[\phi,\gamma]} 
\label{utility}
\end{equation}
where $\lambda$ is a  parameter and  
$M[\phi,\gamma]$ and $R[\phi,\gamma]$ denote the gain and the risk
in the presence of transaction costs, i.e., the expected value and
the variance of
\begin{equation}
\Delta W - Cr^{N} =  \sum_{i=0}^{N-1}(u_i-r) r^{N-i-1} S_i \phi_i(S_i) - \{S_T-S_c\}^+
-\gamma F[\phi]
\label{eq:DeltaW_gamma}
\end{equation}
The gain and the risk are expanded in powers of $\gamma$:
\begin{eqnarray}
M[\phi,\gamma] &=& M_0[\phi] - \gamma M_1[\phi] \\
R[\phi,\gamma] &=& R_0[\phi] + \gamma R_1[\phi] + \gamma^2 R_2[\phi] \\
\end{eqnarray}
where $M_0[\phi]$ is identical to 
(\ref{drift_appendix}) and  $\gamma M_1[\phi]$,
equal to $\gamma <F[\phi]>$,
are the expected trading costs using $\phi$.
We look for strategies that are expandable
in power series in $\gamma$
\begin{equation}
\phi =\phi_0+ \gamma \phi_1 + \ldots 
\end{equation}
To maximize the utility (\ref{utility}) we first wish to minimize
the risk at constant gain.
We minimize
\begin{eqnarray}
Q[\phi,\gamma] &=& R[\phi,\gamma] -2q\left(M[\phi,\gamma] - m\right)
\end{eqnarray}
by varying $\phi$ and $q$. The first step gives
\begin{eqnarray}
\phi(q) &=& \phi_0(q)
-\gamma (\frac{\delta^2 R_0[\phi]}{\delta\phi^2}|_{\phi_0})^{-1}
\cdot [\frac{\delta R_1[\phi]}{\delta \phi}|_{\phi_0}
+2q \frac{\delta M_1[\phi]}{\delta \phi}|_{\phi_0}]
+ {\cal O}(\gamma^2)
\end{eqnarray}
By varying with respect to $q$ we have
\begin{eqnarray}
q(m) & =& q_0(m)  
-\gamma \frac{-M_1[\phi_0(q_0)] + (\frac{\delta M_0[\phi]}{\delta\phi}|_{\phi_0(q_0)})
\cdot\phi_1(q_0(m))}{
(\frac{\delta M_0[\phi]}{\delta\phi}|_{\phi_0(q_0)})
\cdot(\frac{\partial \phi_0(q)}{\partial q}|_{q_0})}
+ {\cal O}(\gamma^2)
\end{eqnarray}
The solutions to the
zeroth order equations are the same as those discussed above 
in section~\ref{s:no-transaction-costs},
\begin{eqnarray}
q_0(m) &=& -\nu(m+\Ach) \qquad
\tilde\psi_i(S_i;q) = (K^{-1}  {\bf F})_{i}(S_i)
 - \mu(\Bach  + q)(K^{-1} {\bf 1})_i(S_i)
\label{Gofq}
\end{eqnarray}
where $\nu$ and $\Ach$ are given in 
(\ref{nudef}) and (\ref{Mhatdef}), and 
$\tilde\psi$ is identical to (\ref{GofM}),
only expressed
as a function of the Lagrange multiplier $q$.
As a function of $m$
the optimal strategy may be written
\begin{eqnarray}
\phi(m) &=& \phi_0(q_0(m)+\gamma q_1(m)+\ldots) +
	\gamma \phi_1(q_0(m)+\gamma q_1(m)+\ldots) +\ldots \nonumber \\
&=& \phi_0(q_0(m)) + \gamma\left(\phi_1(q_0(m)) + 
(\frac{\partial \phi_0(q)}{\partial q}|_{q=q_0(m)}) q_1(m)\right)
 + {\cal O}(\gamma^2)
\end{eqnarray}
which may be simplified to
\begin{eqnarray}
\phi_0(q_0(m))+\gamma
\frac{M_1[\phi_0(q_0(m))]  }
{
(\frac{\delta M_0[\phi]}{\delta\phi}|_{\phi_0(q_0(m))})\cdot
(\frac{\partial \phi_0(q)}{\partial q}|_{q_0(m)})}
\left(\frac{\partial \phi_0(q)}{\partial q}|_{q_0(m)}\right)
+ {\cal O}(\gamma^2)
\label{first-order-strategy}
\end{eqnarray}

The minimal risk as a function of trading gain can now be expressed as
\begin{eqnarray}
R[m] &=& 
R_0[\phi_0(q_0(m)+\gamma q_1(m)+\ldots)
+ \gamma \phi_1(q_0(m)+\gamma q_1(m)+\ldots) + \ldots] \nonumber \\
&& + \gamma R_1[\phi_0(q_0(m)+\gamma q_1(m)+\ldots)
+ \gamma \phi_1(q_0(m)+\gamma q_1(m)+\ldots) + \ldots] + \dots 
\end{eqnarray}
which can be expanded into
\begin{eqnarray}
R[m] &=& R_0[\phi_0(q_0(m))]
		+\gamma \{
		\frac{\delta R_0}{\delta \phi}\cdot
		\left(\frac{\partial\phi_0}{\partial q}\cdot q_1(m) 
		+ \phi_1(q_0(m))\right)
		 \nonumber \\
&&	+R_1[\phi_0(q_0(m))]\} + 
{\cal O}(\gamma^2) 
\end{eqnarray}
Since the combination of $\phi_1(q_0(m))$ and 
$\frac{\partial\phi_0}{\partial q}\cdot q_1(m)$ simplifies
we have, in fact, a much more compact expression for the risk
as a function of $m$,
expanded up to first order in $\gamma$, namely
\begin{eqnarray}
R[m] &=& 
R_0[\phi_0(q_0(m))]
+\gamma \left(\frac{dR_0[m]}{dm}\cdot M_1[\phi_0(q_0(m))]
+R_1[\phi_0(q_0(m))]\right) + {\cal O}(\gamma^2)
\label{eq:with-cancellation}
\end{eqnarray}
The interpretation of (\ref{eq:with-cancellation})
is straight-forward.
The expected trading costs, 
using the zeroth order trading strategy,
appropriate for a level
of gain equal to $m$,
is $\gamma M_1[\phi_0(q_0(m))]$.
When we use this strategy we actually realise
a gain of $m- \gamma M_1[\phi_0(q_0(m))] $.
To reach a level
of $m$, compensating for trading losses, we must use a strategy that in
the absence of trading costs would give
$m +\gamma M_1[\phi_0(q_0(m))] $. In
doing so we run up the extra risk,
to first order, of
$\gamma \frac{dR_0[m]}{dm}\cdot M_1[\phi_0(q_0(m))]$.

Since the risk is now expressed as a function of gain and the
trading cost parameter $\gamma$, we can write
the utility as
\begin{equation}
U[m,\gamma;\lambda] = m+Cr^N -\lambda\sqrt{R[m,\gamma]} 
\label{utility_function_of_m}
\end{equation}
At given risk-aversiveness parameter $\lambda$ we seek to
maximize the utility by varying $m$. Let us assume that the
maximum is obtained at a value of $m[\lambda,\gamma]$. We then observe
that a price acceptable to the writer must be such that the
resulting utility is non-negative. If the option writer does
not sell an option and performs no operations in the market,
then his incremented wealth is identically zero, which carries
zero utility. We hence have
\begin{equation}
C[\gamma;\lambda] = r^{-N}\left(-m[\lambda,\gamma]+\lambda\sqrt{R[m[\lambda,\gamma],\gamma]} \right)
\label{minimum_price}
\end{equation}
where we understand that this is the lowest price that this option writer
is prepared to ask for.

We are therefore to maximize 
the expression $m -\lambda\sqrt{R[m,\gamma]} $,
and we do so by
expanding
\begin{eqnarray}
m[\lambda,\gamma] &=& m_0[\lambda] + \gamma m_1[\lambda] + \ldots
\label{m_expansion}
\end{eqnarray}
and all the preceeding expansions, which together yield
\begin{eqnarray}
U[m,\gamma;\lambda] - r^{N} C & = &
\left(m_0[\lambda]-
\lambda\sqrt{R_0[m_0[\lambda]]}
\right) \nonumber \\
&& +\gamma\{m_1[\lambda]
-\frac{1}{2}\lambda\frac{1}
{\sqrt{R_0[m_0[\lambda]]}}
(\frac{dR_0[m]}{dm}\cdot 
m_1[\lambda] + 
R_1[\phi_0(m_0[\lambda])] \nonumber \\
&&\quad +	\frac{dR_0[m]}{dm} M_1[\phi_0(q_0(m_0[\lambda]))]
			)\} + {\cal O}(\gamma^2)
\label{utility_expanded_in_gamma}
\end{eqnarray}

All zeroth order calculations can be done rather explicitly since
we have $R_0[m] = \rho+\nu(m+\Ach)^2$, with all the coefficients
known. Hence
\begin{eqnarray}
m_0[\lambda] & =& -\Ach + \sqrt{\frac{\rho}{\nu(\lambda^2\nu-1)}} \\
C[\gamma=0;\lambda] & =& r^{-N}\left(\Ach + \sqrt{\frac{\rho(\lambda^2\nu-1)}{\nu}}\right)
\end{eqnarray}
with $\rho$, $\Ach$ and $\nu$ given in
(\ref{rhodef}), (\ref{Mhatdef}) and (\ref{nudef}).
The corresponding (zeroth order) optimal strategy is
\begin{eqnarray}
\tilde\psi_i(S_i;m_0[\lambda]) &=&
\left((K^{-1}{\bf  F})_i(S_i) -\mu\Bach (K^{-1}{\bf 1})_i(S_i)\right) 
+ \sqrt{\frac{\rho\nu}{\lambda^2\nu-1}}(K^{-1}{\bf 1})_i(S_i)
\label{strategy-zeroth-order}
\end{eqnarray}
This is again equivalent to
(\ref{GofM}) and (\ref{Gofq}), but expressed
this time as a function of the risk-aversiveness
parameter $\lambda$.
The structure is that of
a $\lambda$-dependent correction to
the risk-minimizing strategy 
(\ref{GofMstar}).
When $\lambda$ is
large the correction is small.
When $\lambda$ diminshes, such that the
combination $\lambda^2\nu$
tends to one from above, the correction is large.
The case of $\lambda^2\nu$ less than one can be treated by
comparing with pure stock investment\cite{AurellZyczkowski}.
An operator using utility 
function (\ref{utility}) would then invest an unlimited amount in
stock. In other words, when $\lambda^2\nu-1$ is negative, the
formulation of the problem using (\ref{utility}) gives unreasonable
results, both for stock and options. If, however, the utility function is
modified by adding a quantity $-\frac{1}{2W_0}R[\phi]$, then the
stock investor only invests an amount proportional to $W_0$.
The prefactor $W_0$ is hence a measure of the amount of money
an agent can invest in the market. The option price can then be
fixed by comparing the utilities of option trading and pure
stock investment, in a similar way as the option price has here
been fixed by comparing option trading and doing nothing.
The general structure of the solution will again be the
risk-minimizing strategy (\ref{GofMstar}) and the price in
the risk-minimizing approach, with corrections which will now depend
on both $\lambda$ and $W_0$, see \cite{AurellZyczkowski}.
For the rest of this paper we will assume that $\lambda^2\nu$ is
greater than one.

Coming back to (\ref{utility_expanded_in_gamma})
we see that
the structure of $R_0$  also facilitates the optimization
to next order in $\gamma$, since
\begin{eqnarray}
\frac{1}{2}\lambda\frac{1}{\sqrt{R_0[m_0[\lambda]]}}
\cdot\frac{dR_0[m]}{dm}|_{m=m_0[\lambda]}\, =\,1
\end{eqnarray}
We therefore have up to first order in $\gamma$
\begin{eqnarray}
U[m,\gamma;\lambda] - r^{N} C & =& \left(m_0[\lambda]-\lambda\sqrt{R_0[m_0[\lambda]]}
					\right) \nonumber \\
&& -\gamma\left(M_1[\phi_0(m_0[\lambda])]
+
\frac{1}{2}
\sqrt{\frac{\lambda^2\nu-1}{\rho\nu}} R_1[\phi_0(m_0[\lambda])]\right) + {\cal O}(\gamma^2) 
\label{utility_simplified_expansion_in_gamma}
\end{eqnarray}
which does not depend on $m_1[\lambda]$. Hence we do not need to compute
$m_1[\lambda]$.
The price, fixed by the requirement that the maximal utility is zero,
is finally
\begin{eqnarray}
C[\gamma;\lambda]  & =& r^{-N}\left(
\Ach + \sqrt{\frac{\rho(\lambda^2\nu-1)}{\nu}}\right) + \nonumber \\
&& \gamma r^{-N}\left(M_1[\phi_0[\lambda]] +\frac{1}{2}
\sqrt{\frac{\lambda^2\nu-1}{\rho\nu}} R_1[\phi_0(m_0[\lambda])]\right) + {\cal O}(\gamma^2)
\label{price_final}
\end{eqnarray}

Using
(\ref{GofMstar}), (\ref{Gofq}),
(\ref{first-order-strategy}) and
(\ref{strategy-zeroth-order}) we can also express the optimal
strategy up to first order in $\gamma$ as the risk-minimizing strategy
and a correction proportional
to $(K^{-1}{\bf 1})_i(S_i)$.

\section{MARKET EQUILIBRIUM}
\label{s:buyer-seller} 
Rationalizations of market prices are economically meaningful if there
are both buyers and sellers. We have so far exposed the point of
view of the writer of the option.
In the context of mean-variance arguments the analysis for the buyer is 
however very similar.
If, in fact, an option buyer would use a strategy $\phi^{b}$, then,
in the absence of trading costs, 
his incremental wealth would be equal in size
but opposite in sign
of that of an option 
writer using $-\phi^{b}$.
From this follows immediately the strategy and the price appropriate
for an option buyer described by a risk-aversiveness parameter $\lambda$.
The bid-ask spread of agents is thus, in the absence of trading costs,
\begin{eqnarray}
C^{\hbox{bid/ask}}[\gamma=0;\lambda]  & =& r^{-N}\left(
\Ach \pm \sqrt{\frac{\rho(\lambda^2\nu-1)}{\nu}}\right)
\label{bid-ask-spread}
\end{eqnarray}
The only agents prepared to buy and sell are those with a value
of $\lambda$ such that $\lambda^2\nu-1$ vanishes, and the price
they offer is $r^{-N}\Ach$, which we recognize by
(\ref{Cdefinition}) to be the same as that from the risk-minimizing 
procedure.

The trading costs break the symmetry between buyer and writer. All things being
equal, the buyer will be prepared to pay a little less, and the writer
will ask a little more. A gap opens between the
smallest ask price and the largest bid price. Strictly speaking, we
do not find a market price in the
presence of trading costs.
When $\lambda^2\nu$ tends to one from above,
the zeroth order strategy,
$\phi_0(m[\lambda])$, diverges. Hence the average
trading costs become large.
It is convenient to introduce the notation
\begin{eqnarray}
\phi^{\hbox{risk}} &=& \frac{1}{P(S_i|S_0)
 r^{N-i-1}S_i}\, 
\left((K^{-1}{\bf  F})_i(S_i) -\mu\Bach (K^{-1}{\bf 1})_i(S_i)\right) \\
\phi^{\hbox{gain}} &=& \frac{1}{P(S_i|S_0)
 r^{N-i-1}S_i}\, (K^{-1}{\bf 1})_i(S_i)
\label{buy-and-hold-strategy}
\end{eqnarray}
such that equation (\ref{strategy-zeroth-order}) can be written for
the variable $\phi$
as
\begin{eqnarray}
\phi_0(\lambda) &=& \phi^{\hbox{risk}}
+ \sqrt{\frac{\rho\nu}{\lambda^2\nu-1}}
\,\phi^{\hbox{gain}}.
\label{phi-0-strategy-function-of-lambda}
\end{eqnarray}
The first order correction to the strategy can by
(\ref{Gofq}) and
(\ref{first-order-strategy}) be written
\begin{eqnarray}
\phi_1(\lambda) &=& \mu\nu M_1[\phi_0(\lambda)]\,\phi^{\hbox{gain}}.
\label{phi-1-strategy-function-of-lambda}
\end{eqnarray}
When $\lambda^2\nu$ is close to one the 
trading costs are dominated by the component
proportional to $\phi^{\hbox{gain}}$.
The only way in which the dominating contribution
in this case
could come from $\phi^{\hbox{risk}}$ would be if
$\phi^{\hbox{gain}}$ were actually a
buy-and-hold strategy.
It is fairly straight-forward to see that
this is not so. It
suffices to look at the case when $\mu$ is
equal to zero, the matrix
$K$ diagonal, and use
(\ref{K-diagonal-definition}).

When $\lambda^2\nu$ is larger, $\phi^{\hbox{gain}}$ and $\phi^{\hbox{risk}}$
contribute each to the trading costs.
We thus have 
\begin{eqnarray}
M_1[\phi_0(\lambda)] =
\sqrt{\frac{\rho\nu}{\lambda^2\nu-1}}
M_1[\phi^{\hbox{gain}}]
\,+\, M_1[\phi^{\hbox{risk}}]
\label{trading-costs-in-gamma}
\end{eqnarray}
We can also address in a similar manner the
first order
correction to the risk that enters
in (\ref{price_final}).
When $\lambda^2\nu$ is close to one we should
have
\begin{eqnarray}
\sqrt{\frac{\lambda^2\nu-1}{\rho\nu}}
R_1[\phi_0(\lambda)] \sim
\sqrt{\frac{\rho\nu}{\lambda^2\nu-1}}
R_1[\phi^{\hbox{gain}}]
\end{eqnarray}
which depends in a similar way as the first term
in (\ref{trading-costs-in-gamma}) on $\lambda$.
After a possible redefinition of
$M_1[[\phi^{\hbox{gain}}]$
to incorporate the
incremental risk $R_1$ we can
therefore write the bid/ask
prices as
\begin{eqnarray}
C^{\hbox{bid/ask}}[\gamma;\lambda]  & =&
r^{-N}\left(
\Ach \pm
\left(\sqrt{\frac{\rho(\lambda^2\nu-1)}{\nu}} 
+\gamma M_1[\phi^{\hbox{gain}}]
\sqrt{\frac{\rho\nu}{\lambda^2\nu-1}} +\gamma M_1[\phi^{\hbox{risk}}]  \right).
\right)
\label{bid-ask-spread-with-gamma}
\end{eqnarray}

We can now find the optimal $\lambda$ as a
function of $\gamma$ by minimizing the bid-ask
spread. The minimum is attained at
$(\lambda^*)^2 = \frac{1}{\nu}
+\gamma 
M_1[\phi^{\hbox{gain}}]$
and leads to
\begin{eqnarray}
C^{\hbox{bid/ask}}[\gamma;\lambda^*] & = &
r^{-N}\left( \Ach \pm\left(
2\sqrt{\rho\gamma M_1[\phi^{\hbox{gain}}]} +\gamma M_1[\phi^{\hbox{risk}}]
\right)\right)
\label{bid-ask-spread-with-gamma-star}
\end{eqnarray}
The two components in the bid-ask spread are the expected trading costs
using the risk-minimizing strategy, $\gamma M_1[\phi^{\hbox{risk}}]$,
and a term proportional to $\sqrt{\rho}$, the square root of the minimal
residual risk. This is also the form used in \cite{BouchaudSornette}.
The new piece in (\ref{bid-ask-spread-with-gamma-star}) is that the
proportionality factor of $\sqrt{\rho}$ is not a free parameter, but
in principle computable. Note that 
$\phi^{\hbox{gain}}$ does not involve the pay-out function of the option.
Hence $\gamma M_1[\phi^{\hbox{gain}}]$ takes the same value for different options.
The optimal stategies are approximately
\begin{eqnarray}
\phi^{\hbox{writer/buyer}}(\lambda^*)
= \pm \phi^{\hbox{risk}}
+ 
\sqrt{\frac{\rho}{\gamma
M_1[\phi^{\hbox{gain}}]}}
\,\phi^{\hbox{gain}}
\label{writer-buyer-strategies}
\end{eqnarray}

In a mean-variance world with transaction costs
and homogeneous expectations there is a minimum
gap between the smallest ask and the largest
bid price. We should therefore, in that
world, not expect to see any trading in options
at all.
In the real world
all agents do not of course hold identical
expectations. In addition, there is
no reason to expect that all agents 
are necessarily
well described by their preferred trade-offs
of risk vs. return, measured by a variance and
an expected value.
In somewhat extreme parameter ranges 
mean-variance option pricing lead to paradoxical
results
\cite{Wolczynska,Hammarlid,AurellSimdyankin},
also discussed 
in the finance literature
quite some time ago \cite{DybvigIngersoll}.

It is therefore not a problem {\it per se} that
we find a gap between bid and ask prices.
The price $\Ach$ from
(\ref{bid-ask-spread-with-gamma-star}) and
(\ref{Mhatdef}) is an estimate of a market
price, and it is in the end
an empirical question to decide
how useful that estimate is. 
If the gap as predicted by
(\ref{bid-ask-spread-with-gamma-star})
is sufficiently small compared to inhomogeneous expectations
and all other externalities that 
tend to move and push apart
bids and asks, then there is no conceptual difference between
testing  $\Ach$ or the Black-Scholes price against observed
market data. For tests of the risk-minimizing prescriptions
against market data, chiefly for the spacial case $\mu=0$,
see \cite{BouchaudSornette,BouchaudPotters}.

The analysis of this paper has been
performed in perturbation theory in $\gamma$.
This assumes that the trading costs are relatively small.
It is however perfectly natural to
consider the case where the minimal risk $\rho$
is very small, but trading with the strategy
$\phi^{\hbox{risk}}$ leads
to very large costs. For instance, suppose that risk is minimized
by rehedging very often, and one pays some amount for every trade,
an example considered in\cite{BouchaudSornette}.
It then seems clear that the best strategy is probably not
very close to the optimal strategy without transaction costs, and
the perturbation cannot really be considered small. 
We may however get information also on this case by imagining
that the space of possible strategies may be varied
(for instance, by trading more or less often).
The minimal risk and the trading costs will then both
change with the class considered. 
In minimizing the bid-ask spread there is
a trade-off between
minimizing minimal risk (by trading more often)
and minimizing trading costs (by trading less
often). One expects that the best trade-off is
obtained when minimal risk and trading costs
are of the same order.

To conclude, the mean-variance approach to option pricing leads to
the same price as the risk-minimizing prescription of~\cite{BouchaudSornette}
and~\cite{Schweizer}. The optimal strategy is, in the absence of transaction costs,
equal to the risk-minimizing strategy plus another strategy more related to
direct investment in stock. Transaction costs lead to a gap between bid and
ask prices. By minimizing this gap one arrives at a specific level of risk-taking
appropriate to a given level of transaction costs.
All these calculations can be done perturbatively around the case of no
trading friction, and therefore assume that transaction costs using the optimal (zero-cost)
strategy are small. The case when the transaction costs using the optimal (zero-cost)
strategy are actually large can be done in a somewhat more heuristic manner by varying the
class of strategies considered. This leads to the
estimate that the smallest bid/ask spreads are
obtained when trading costs and residual risk
are about
equal. 

The results of the mean-variance approach therefore
finally agree in suprising detail with those of the 
risk-minimizing prescription.

\section{ACKNOWLEDGEMENTS}
We thank Jean-Philippe Bouchaud, Ola Hammarlid,
Sergey Simdyankin and Gra\.zyna Wolczynska for
discussions. E.A. thanks the organizers
of the conference ``Disorder and Chaos''
(Rome September 22-24, 1997) for an invitation.
This work was supported by the Swedish Natural
Science Research Council through grant
S-FO 1778-302 (E.A.),  and by
the Polish State Committee for Research (K.\.Z.).

\end{document}